\documentclass[aps,prd,twocolumn,nofootinbib,superscriptaddress]{revtex4-2}

\usepackage{graphicx,amssymb,amsmath,amsthm,amsfonts,epsfig,epsf,fixmath}
\usepackage{bm}
\usepackage{dcolumn}
\usepackage{latexsym}
\usepackage{rotating}
\usepackage{longtable}
\usepackage{enumerate}
\usepackage{tensor,multirow}
\usepackage{url}
\usepackage{enumitem}
\usepackage{comment}
\usepackage[usenames]{color}
\usepackage{epstopdf}
\usepackage{tikz,tikz-3dplot}
\usepackage{aas_macros}
\usepackage{ulem}
\usepackage{slashed}
\usepackage[scr=boondox]{mathalfa}
\usepackage[scr=boondox]{mathalfa}
\usepackage{slashed}
\usepackage{ulem}
\usepackage{tikz,tikz-3dplot}
\usepackage{url}
\usepackage{xurl}
\usepackage{mathtools}

\DeclareFontFamily{U}{min}{}
\DeclareFontShape{U}{min}{m}{n}{<-> udmj30}{}

\usepackage[colorlinks,citecolor=blue,urlcolor=magenta,linkcolor=blue,linktocpage,breaklinks=true]{hyperref}
\def\sect#1{{Sec.~\ref{#1}}}

\def\sect#1{{Sec.~\ref{#1}}}

\labelformat{subsection}{Sec. #1} 
\labelformat{equation}{Eq.~(#1)} 
\labelformat{figure}{Fig.~#1} 
\labelformat{subfigure}{Fig.~(\thefigure#1)} 
\labelformat{table}{Tab.~(#1)} 
\def\sect#1{{Sec.~\ref{#1}}}

\def\sect#1{{Sec.~\ref{#1}}}

\begin{document}
%%%%%%%%%%%%%%%%%%%%%%%%%%%%%%%%%%%%%%%%%%%%%%%%%%%%%%%%%%%%%%%%%%%%%%%%%%
\title{Dark matter affecting binary evolution: formalism for generic orbits}

\author{Shrijaya Maity}
\affiliation{Department of Space, Planetary and Astronomical Sciences and Engineering (SPASE), Indian Institute of Technology Kanpur, Kanpur 208016, India}
\affiliation{Department of Physics, Birla Institute of Technology and Science Pilani, Rajasthan 333031, India}

\author{Sajal Mukherjee}
 \email{sajal.mukherjee@pilani.bits-pilani.ac.in}
 \affiliation{Department of Physics, Birla Institute of Technology and Science Pilani, Rajasthan 333031, India}

\author{Ranjan Laha}
\affiliation{ Centre for High Energy Physics, Indian Institute of Science, Bengaluru, India}
%%%
%%%
\begin{abstract}
In the present paper, we investigate the evolution of a compact object orbiting a slowly rotating massive body surrounded by an anisotropic dark matter distribution. 
The dynamical friction arising from the surrounding dark matter dissipates the orbital energy, momentum, and Carter constant, modifying the evolution of generic eccentric and inclined orbits. 
Within the linear-spin approximation of the central object, we derive the evolution of these orbits under the combined effects of gravitational radiation and dark matter dissipation. We find that the inclination evolves because of the spin of the central object, and that the angular anisotropy of the dark matter profile changes the cumulative dephasing of the emitted waveform. In addition, we compare the waveform and dephasing with and without the anisotropy. We believe that the present framework can be particularly useful to incorporate anisotropic environmental effects, which are relevant for modeling of generic orbits in long-lived inspiral -- typically occurring in extreme and intermediate-mass-ratio systems.
\end{abstract}
\maketitle
\section{Introduction}
%%%%
%
The evolution of compact object binaries provides a useful perspective to probe both fundamental gravity and the astrophysical environment surrounding them \cite{Barausse:2014tra, Figueiredo_2023,Cardoso:2022whc,Vicente:2022ivh,Yue_2019}. These astrophysical environments can be treated as perturbative effects, and typically affect binary evolution\,\cite{Eda_2015,1973ApJ...185..635T}. A few examples of these environmental effects would include the presence of an external tidal field \cite{PhysRevD.80.084018, PhysRevD.101.064028,lima2022tidal}, third body \cite{Ford_2000,valtonen2006three}, accretion \cite{cardoso2020constraints, kremer2017accreting,frank2002accretion,Radulea:2026hub}, metric deformation \cite{babichev2024testing,chua2018towards}, or the presence of dark matter \cite{PhysRevLett.110.221101}. The last one is the topic of the present paper, particularly the dark matter spikes \cite{herrera2026formationredshiftevolutiondark,Mitra:2025}. 

It has been known for a long time that dark matter forms spikes around massive black holes (BHs) \cite{gondolo1999dark}. The formation of these spikes is well studied in the context of non-rotating static spacetime \cite{lacroix2018dynamical}. Moreover, the implications of these spikes in gravitational wave (GW) astronomy are also not new, and explored extensively in recent times \cite{Becker:2021ivq, Coogan:2021uqv}. This is particularly important as the interactions between the binary constituents and surrounding dark matter can introduce dissipative forces, which modify the inspiral timescale and phase calculations \cite{Speeney:2022ryg}. Note that these dissipative effects appear because of the drag force produced by the dark matter distribution, which causes changes in binary parameters, such as eccentricity, semi-major axis, and inclination.

However, most of these earlier studies are concerned with studying these effects in non-rotating static spacetime, which essentially indicates that the dark matter distribution is isotropic. Therefore, without losing any generality, one may consider $\theta=\pi/2$ plane, and that remains constant throughout the evolution. In other words, the inclination is not affected by either the GW radiation or the dark matter dissipation. In the present work, we relax this condition and allow the spacetime to have rotation and the dark matter distribution to be anisotropic. As expected, the rotation of the spacetime introduces an evolution of the inclination, which is further impacted by the dark matter distribution. This would allow us to model the binary in a generic sense.

With this, we should now touch upon the binary model that we have in mind. Dark matter spikes are expected to form around intermediate-mass black holes ($\sim10^{2}$–$10^{4}\,M_\odot$) \cite{Rose_2022, PhysRevLett.110.221101} and around supermassive black holes \cite{Speeney:2022ryg}. Both are well suited to hosting binaries with a small mass ratio, widely known as intermediate/extreme mass ratio inspirals (I/EMRIs) \cite{Barack:2003fp,Amaro_Seoane_2007}. They have a longer evolution, and therefore, they are sensitive to environments around them \cite{Cole:2022yzw}. These perturbative effects are negligible for a single cycle. However, as the evolution consists of several cycles, their effects gradually grow over time and become significant \cite{Yue_2019}. Our primary objective is to study an I/EMRI in which the central black hole contains an anisotropic dark matter distribution around it. 

For a generic binary evolution, all three constants of motion, namely, energy, angular momentum, and Carter constant, become important \cite{Ryan:1995zm,Ori:1997be,Flanagan:2007tv}. As the binary evolves in time under dissipation, the change of these quantities would relate to the change in eccentricity, inclination, and semi-major axis \cite{Hughes:1999bq}. Within the present framework, we aim to include the effect of both GW radiation and dark matter undergoing a generic evolution. However, to keep the study analytically manageable, we adopt a relatively simpler binary model \cite{PhysRevD.53.3064}, given as follows:
%%%
\begin{itemize}[itemsep=1pt, topsep=4pt, parsep=0pt, partopsep=0pt]
\item The central object of mass $M$ is slowly rotating, and we only consider effects that appear linear in spin.  
\item The secondary object of mass $m$, that is orbiting the central object, is assumed to be non-spinning.
\item The dark matter particles are collisionless and form a spike close to the central object.
\item The gravitational radiation is quadrupolar, while the dark matter dissipative effects are Newtonian. Both consider the leading-order spin effects. 
%
%\item We use the weak-field approximation to address GW radiation, whereas the dark matter results are purely Newtonian.
\end{itemize}
%%%
In addition to the above, note that the anisotropic model considered here represents a \textit{toy model} which is motivated by earlier studies where dark matter distribution is calculated numerically around rotating black holes \cite{Ferrer:2017xwm}. Our primary motivation is to isolate and qualitatively understand the impact of angular anisotropy on the orbital evolution, while restricting ourselves to this toy model. With a given distribution, we obtain the average change of energy, angular momentum, and Carter constant, and finally, the change in orbital elements, like eccentricity and inclination. Once the orbital evolution is known, we can obtain the dephasing and gravitational waveform.

The rest of the paper is organized as follows. In \sect{sec: Basic framework}, we present the basic framework and the dissipation due to dark matter. In \sect{sec:Evolution}, we compute the orbit-averaged fluxes. In \sect{sec:Numerical Result}, we introduce the dark matter distribution, obtain the evolution of the orbital elements, and present numerical results for the orbital evolution, the waveform and the dephasing. Finally, in \sect{sec: Discussion}, we conclude with a discussion. The appendices contain the equations of motion, the derivation of the dark matter dissipation, the orbital frequencies and phase, and the gravitational-wave fluxes. 
%
%%%%%%%%
\section{Basic framework}
\label{sec: Basic framework}
%%%%%%%%
In this section, we develop the mathematical framework to study the binary evolution within an anisotropic dark matter distribution. We adopt the framework presented in Ref. \cite{PhysRevD.53.3064}. The equations of motion are introduced in Appendix \ref{appendix:EOM}, where the Lagrangian is given along with the radial, angular, and azimuthal equations. Following Ref. \cite{PhysRevD.53.3064}, we can express the orbital elements in terms of energy ($\mathcal{E}$), angular momentum ($L_{\rm z}$), and Carter constant ($K$):  
%
%%%
\begin{eqnarray}
&& e^2 = \left[ 1 + \dfrac{2 \mathcal{E} {K}}{m^3 M^2}  \left( 1 + \frac{4 S M m^3 L_{\rm z}}{{K}^2} \right) \right], \nonumber \\
&& a = -\dfrac{M m}{2 \mathcal{E}} \left( 1 + \frac{2 S M m^3 L_{\rm z}}{{K}^2} \right),~\cos \iota = L_{\rm z}{K}^{-1/2}. \nonumber \\
\label{D1}
\end{eqnarray}
%%%%%
In the above expression, notations $e$, $a$ and $\iota$ are eccentricity, semi-major axis, and inclination angle, respectively. The spin of the Kerr spacetime is defined as $S$, and we ignore terms $\mathcal{O}(S^2)$ and beyond. For a limiting case of equatorial plane, $\iota=0$, and ${K}=L^2_{\rm z}$ \cite{PhysRevD.53.3064}. It is also possible to revert the above expressions and obtain the constants in terms of orbital parameters -- 
%%%
\begin{eqnarray}
&& \mathcal{E} = -\dfrac{mM}{2a}-\dfrac{m^4M^2 S \cos\iota}{a\Big(m^2Ma(1-e^2)\Big)^{3/2}}, \nonumber \\
&& L_{\rm z} = \sqrt{Mam^2 (1-e^2)}\cos\iota-\dfrac{3mS \cos^2\iota}{a(1-e^2)}, \nonumber \\
&& {K} = Mm^2a~(1-e^2)-\dfrac{6Mm^3S \cos\iota}{\sqrt{Mm^2a(1-e^2)}}. \nonumber \\
\label{Constant_to_parameter}
\end{eqnarray}
%%%%%
Interestingly, in the above expressions, $\mathcal{E}$ is not the Killing energy $E=-\xi^{\mu}P_{\mu}$, where $P^{\mu}$ is the 4-momentum. The quantity $\mathcal{E}$ is expressed as the Newtonian orbital energy or the binding energy of the binary. The relation between $E$ and $\mathcal{E}$ is given as $ E=m+\mathcal{E}$, and we will be using this relation whenever required. 

Let us now discuss how the presence of a dark matter distribution, $\rho_{dm}(r,\theta)$, introduces dissipative effects on the binary evolution \cite{1943ApJ....97..255C, dosopoulou2024dynamicalfrictiondarkmatter,Chandrasekhar1949}. The detailed calculations are relegated to the Appendix~\ref{appendix:dissipation_dm}, and here we write down the final expressions only:
%%%
\begin{eqnarray}
  &&  \dot{E}=\dfrac{dE}{dt}=-\dfrac{\beta \eta^2}{v} \bar{\rho}_{dm}(r,\theta),  \nonumber \\
  &&  \dot{L_z}=\dfrac{dL_z}{dt} = -\dfrac{\beta \eta^2 \bar{\rho}_{dm}(r,\theta)}{mv^3}\Big(L_z+\dfrac{2mS \sin^2\theta}{r}\Big), \nonumber \\
  &&   \dot{K}= \dfrac{dK}{dt} =-2 \dfrac{\beta \eta^2 \bar{\rho}_{dm}(r,\theta)}{mv^3}K-\dfrac{4S \beta \eta^2 \bar{\rho}_{dm}(r,\theta)}{r v^3}L_{\rm z}. \nonumber \\
  \label{eq:dissip}
\end{eqnarray}
%%%
In the above, $\beta=4 \pi \ln \sqrt{M/m}$, $v$ is the velocity of the secondary object of mass $m$, mass ratio $\eta=m/M$, and  $\bar{\rho}_{dm}(r,\theta)$ is the normalized dark matter distribution, i.e., $\bar{\rho}_{dm}(r,\theta)=M^2\rho_{dm}(r,\theta)$ \cite{Kavanagh_2020}. With these expressions, $\dot{E}$, $\dot{L_z}$, and $\dot{K}$, we now attempt to relate them to the change of orbital elements, i.e., $e(t)$, $a(t)$, and $\iota(t)$. Within this description, at every instant, the orbit is specified by the instantaneous values of $({E},L_z,K)$, and the drag changes these quantities slowly. The rates in \ref{eq:dissip} are the changes produced by the force at a fixed position. Using \ref{D1} and the chain rule, these instantaneous rates give $(\dot a,\dot e,\dot\iota)$. Since the orbital elements change little over one orbit, we average over a radial period (\sect{sec:Evolution}) to obtain the secular evolution.

In passing, we should emphasize that the drag force considered here (details are given in Appendix~\ref{appendix:dissipation_dm}) is the Newtonian dynamical-friction expression, while the orbital dynamics keeps the linear-in-spin corrections. We do not include relativistic corrections to the drag, so this is not a fully post-Newtonian treatment. The spin terms are kept because they alone produce the inclination evolution. The Newtonian drag is therefore a reference model, and relativistic corrections to the drag are left for future work.
%%%%%%%%%%%%%%%%%%%%%%%%%%
\section{Calculating time-averaged flux}
\label{sec:Evolution}
%%%
To obtain the fluxes, we start with the following reparameterization of the orbits \cite{PhysRevD.53.3064}:
%%%%%
\begin{eqnarray}
 r(\psi)&=& \dfrac{K/(m^2 M)}{1+e \cos(\psi)}\Big(1+\dfrac{SL_z m^3 M}{K^2} \big[6+2e \cos(\psi)\big]\Big),\nonumber \\
 \cos\theta &=&\sin \iota\sin(\psi),
\label{eq:r_psi_th_chi}
\end{eqnarray}
%%%%%
where $0 \leq \psi \leq 2\pi$. Note that $\psi$ is used to parametrize both $r$ and $\theta$, which is not valid for generic Kerr orbits. However, as the present formalism is applicable within the linear-in-spin and weak-field approximations, we will use the above to obtain the orbit-averaged fluxes:  
\begin{equation}
   \Big \langle  \dfrac{dE}{dt} \Big\rangle_{t}=\dfrac{1}{T_r}\int^{T_r}_{0} \dfrac{dE}{dt} dt,
\end{equation}
where $\langle \rangle$ stands for averaging, and $T_r=2\pi/\Omega_r$ is the radial time-period, where $\Omega_r$ is given in Appendix~\ref{sec:dephasing}. 
% 
%%%%
\subsection{General case}
%%%
In this case, the dark matter distribution is kept as $\bar{\rho}_{dm}(r,\theta)$, and we assume no prior form for $\bar{\rho}_{dm}(r,\theta)$. We keep $\bar{\rho}_{dm}$ fixed, i.e.\ no feedback of the companion on the dark matter distribution is included \cite{Kavanagh_2020,Coogan:2021uqv}. With this, the average dissipation \cite{BHPToolkit} would take the form
%%%
%%%
\begin{eqnarray}
    \Big\langle \dfrac{dE}{dt}\Big\rangle_{t} &=&-\beta \eta^2 \Big\langle \dfrac{\bar{\rho}_{dm}(r,\theta)}{v(r)}\Big \rangle_{t}.
\end{eqnarray}
%%%
We can now use the reparameterization given in \ref{eq:r_psi_th_chi}, and obtain
%%%
\begin{eqnarray}
    \Big\langle \dfrac{dE}{dt}\Big\rangle_{t} &=&-\dfrac{\beta \eta^2}{T_r} \int^{2 \pi}_{0} \dfrac{\bar{\rho}_{dm}(r(\psi),\theta(\psi))}{v(r(\psi))} \dfrac{dt}{d\psi}d\psi.\nonumber \\
\end{eqnarray}
%%%
Now, within the present context, we have both $r$ and $\theta$ parametrized with $\psi$. Therefore, the above expression can be obtained even if both $r$ and $\theta$ are coupled in $\bar{\rho}_{dm}(r,\theta)$. If we relax this condition that both $r$ and $\theta$ are parametrized by $\psi$, and consider an arbitrary spin value, the scenario would be very different, and the above integral can be non-trivial. Keeping that in mind, and also the fact that our motivation is primarily to keep things at the analytical level, we consider a special case where $\rho_{dm}(r,\theta)$ is separable in $r$ and $\theta$. This case is discussed below. 
%
%%%
\subsection{Special case}
%%%
%
Let us now consider the case when dark matter distribution is given by $\bar{\rho}_{dm}(r,\theta)=f(r) g(\theta)$, where the explicit form of $f(r)$ and $g(\theta)$ is not known at this stage. By employing this, we now attempt to obtain the average dissipation due to dark matter. The instantaneous rates are derived in Appendix~\ref{appendix:dissipation_dm}, and the orbit averages are obtained following Ref. \cite{PhysRevD.53.3064}:
%%%
  \begin{eqnarray}
  \Big\langle \dfrac{dE}{dt}\Big\rangle_{t} &=&-\beta \dfrac{m^2}{M^2} \Big\langle \dfrac{f(r)}{v(r)} g(\theta)\Big \rangle_{t}, 
  \label{eq:dE/dlambda_ave}
\end{eqnarray}
%%%
where we exploit the separability between $r$ and $\theta$ in $\bar{\rho}_{dm}(r,\theta)$. 
%
%%%
Similar to the above expression, we can also obtain the average dissipation of momentum as follows:
%%%
\begin{align}
 {\Big \langle \dfrac{dL_z}{dt}\Big\rangle}_{t}=-\frac{\beta m^{2}}{M}\Big[\frac{L_z}{mM}\Big\langle\frac{f(r)\,g(\theta)}{v^{3}}\Big\rangle_t \nonumber \\
+2\frac{S}{M^{2}}\Big\langle\frac{f(r)\,g(\theta)\sin^{2}\theta}{(r/M)\,v^{3}}\Big\rangle_t\Big]
\end{align}
%%%
%%
Finally, the rate of change of the Carter constant can be written as
\begin{eqnarray}
    \dot{{K}} &=& -2 \beta m^3 \dfrac{f(r)g(\theta)}{v^3}\Big(\dfrac{K}{m^2 M^2}\Big) \nonumber \\
    && -{4\beta m^3 (S/M^2)}\dfrac{Mf(r)g(\theta)}{v^3 r}\Big(\dfrac{L_{\rm z}}{m M}\Big)
\end{eqnarray}
%%
%%%
Similar to $L_z$, here we can write (by using \ref{Constant_to_parameter})
%%%
%%%
\begin{widetext}
    \begin{eqnarray}
 {\Big \langle \dfrac{d {K}}{dt}\Big\rangle}_{t}&=& -2\beta m^3 \Big(\dfrac{a(1-e^2)}{M}-\dfrac{6(S/M^2)\cos\iota}{\sqrt{a(1-e^2)/M}}\Big)\Big\langle\dfrac{f(r)g(\theta)}{v^3} \Big \rangle_t-{4\beta m^3 (S/M^2)}\Big([a(1-e^2)/M]^{1/2}\cos\iota\Big)\Big \langle \dfrac{f(r)g(\theta)}{v^3 (r/M)}  \Big \rangle_t. \nonumber \\
 \label{eq:CARTER_Time}
\end{eqnarray}
\end{widetext}
%%%
%%%
Therefore, given an expression for $\bar{\rho}_{dm}(r,\theta)$, we can simply use that in the above relations and obtain the fluxes. In the discussion given below, we have provided one such example. 
%%%%%%%%%%%%%%%%%%%%%%%
\section{Binary evolution with dark matter effects}
\label{sec:Numerical Result}
%%%%%%%%%%%%%%%%%%%%%%%
In the present section, we will discuss an example in which we evolve a binary system undergoing both GW and dark matter dissipation. We first discuss a specific dark matter distribution, and then obtain the fluxes. By using that, we calculate the waveform and dephasing for the emitted GWs. 
%%%%%%%%%%%%%%%%%%%%%%%
\subsection{Dark matter distribution}
%%%%%
Although it is necessary to obtain the $\theta$-dependent distribution from a robust mechanism, in the present paper, we are particularly interested in investigating its effect on the evolution. Therefore, we obtain the $\theta$-dependency from an intuitive viewpoint, guided by the numerical results discussed in Ref. \cite{Ferrer:2017xwm}. As we consider the dark matter distribution, which is decoupled in $r$ and $\theta$, the radial distribution is given by:
%%%%
\begin{equation}
f(r)=
\begin{dcases}
k_1\left(\dfrac{M}{r}\right)^\alpha, & r_{\rm in}<r<r_{\rm sp},\\
0, & r<r_{\rm in}.
\label{eq:dark_matter_density}
\end{dcases}
\end{equation}
%%%%
where $k_1=\bar{\rho}_{\rm sp}(r_{\rm sp}/M)^{\alpha}$, and $r_{\rm in}$ is the inner radius of the spike, which we set at $r_{\rm in}=4M$. Additionally, $r_{\rm sp}$ is the size of the spike, and $\bar{\rho}_{\rm sp}=\rho_{\rm sp}M^2$ is the normalized value of the spike's density at $r_{\rm sp}$. In the present context, we set $\alpha=7/3$ and $k_1 = 10^{-6}$ throughout the paper. The orbit averages below assume that the orbit lies entirely inside the spike, $r_{\rm in}<r_{\rm p}=a(1-e)$ and $r_{\rm a}=a(1+e)<r_{\rm sp}$, which holds for all the orbits we consider.

\begin{figure}[htp]
  \centering  \includegraphics[scale=0.75]{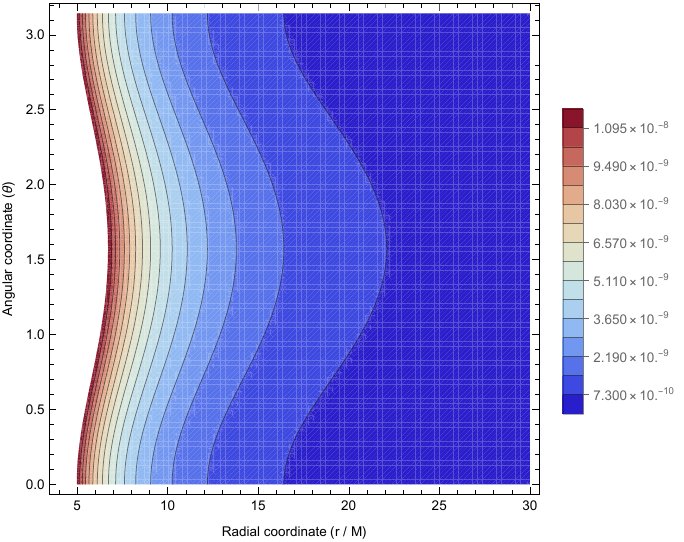}
  \caption{In the above, we illustrate $\bar{\rho}(r,\theta)$ for $k_1=10^{-6}$, and $k_2=1/2$. The color bar on the right displays the numerical value of $\bar{\rho}(r,\theta)$. The density is maximum at the equator, and as we move away from that, the density decreases. In addition, for a given value of $\theta$, as we move away from the horizon, the density decreases.}
  \label{Darkmatterprofile}
\end{figure}

Let us now introduce the $\theta$-dependent part of the distribution. Since the Kerr spacetime is symmetric about the equatorial plane, we expect the dark matter distribution to respect this symmetry. With this logic, $\theta$-dependence can be expanded in even-order Legendre polynomials $P_{2n}(\cos\theta)$,
%%%
\begin{equation}
    \bar{\rho}(r,\theta)=\bar{\rho}_0(r)+\bar{\rho}_2(r)P_2(\cos\theta)+\bar{\rho}_4(r)P_4(\cos\theta)+....
\end{equation}
%%%
and up to the leading order, we rewrite it as:   
\begin{equation}
    \bar{\rho}(r,\theta)=k_1\left(\dfrac{M}{r}\right)^{7/3}\Big(1-k_2 \Big[\dfrac{1+2 P_2(\cos\theta)}{3}\Big]\Big),
        \label{eq:dark_matter_density_theta}
\end{equation}
where we have ignored terms beyond $P_2(\cos\theta)$. In the above expression, $k_2$ behaves as an expansion parameter, and it measures the shape of the distribution. At the equator, the $\theta$-dependency would identically vanish. To maximize $\bar{\rho}(r,\theta)$ at the equator, we must have $k_2>0$ \cite{Ferrer:2017xwm}. At the poles, the value of $\bar{\rho}(r,\theta)$ explicitly depends on $k_2$. Interestingly, for values of $k_2$ exceeding unity, $\bar{\rho}(r,\theta)$ becomes negative at the poles. Therefore, to ensure the distribution is positive everywhere, we consider $0<k_2<1$. The distribution is given in \ref{Darkmatterprofile} for a set of distribution parameters. Note that we do not attempt a quantitative fit of $k_2$ to the numerical profiles of Ref.~\cite{Ferrer:2017xwm}, and we treat $k_2$ as a free parameter that controls the strength of the anisotropy. The fluxes in \ref{eq:K's} and the rates in \ref{eq:evolution} are linear in $k_2$. So, the anisotropic contribution for any other value of $k_2$ can be obtained by rescaling.
%%%%
\vspace{1 cm}

With the dark matter distribution given in \ref{eq:dark_matter_density_theta}, the anisotropy implies that a test particle on an equatorial orbit experiences a different dynamical friction than one on an inclined orbit. As a result, the orbit-averaged rates $\dot{E}$, $\dot{L}_z$ and $\dot{K}$ acquire corrections that modify the phase of the emitted gravitational waves. We obtain them by inserting the orbit parametrization of \ref{eq:r_psi_th_chi} and the density of \ref{eq:dark_matter_density_theta} into the averages of \sect{sec:Evolution}, and expanding up to $\mathcal{O}(e^4,\iota^4,S)$. These are given as:
%%%%%
%%%%%
%%%%
\begin{align}
    \langle\dot{E}_{\mathrm{dm}}\rangle &= -\frac{k_1\,\beta\,\eta^{2}}{(a/M)^{11/6}}
        \Big(\mathcal{E}_1+\frac{S}{M^2}(a/M)^{-3/2}\,\mathcal{E}_2\Big), \nonumber \\[6pt]
    \langle\dot{L_{z}}_{\mathrm{dm}}\rangle &= -\frac{k_1\,\beta\,m^{2}}{M\,(a/M)^{1/3}}
        \Big(\mathcal{L}_1+\frac{S}{M^2}(a/M)^{-3/2}\,\mathcal{L}_2\Big), \nonumber \\[6pt]
    \langle\dot{K}_{\mathrm{dm}}\rangle &= -\,2\,k_1\,\beta\,m^{3}\,(a/M)^{1/6}
        \Big(\mathcal{K}_1-\frac{S}{M^2}(a/M)^{-3/2}\,\mathcal{K}_2\Big).
    \label{eq:ELKdotdm}
\end{align}
%%%%
where,
\begin{widetext}
  \begin{align}
  & \mathcal{E}_1=\left(1+\frac{13}{36}\,e^{2}+\frac{1225}{5184}\,e^{4}\right)
+ k_2\,\iota^{2}\left(\iota^{2}-3\right)
\left(\frac{1}{6}+\frac{7}{432}\,e^{2}+\frac{2005}{93312}\,e^{4}\right), \nonumber \\
 & \mathcal{E}_2= \left(1+\frac{29}{12}\,e^{2}+\frac{18977}{5184}\,e^{4}\right)
\left(1-\frac{\iota^{2}}{2}+\frac{\iota^{4}}{24}\right)
- k_2\,\iota^{2}\left(6-5\iota^{2}\right)
\left(\frac{1}{12}+\frac{7}{96}\,e^{2}+\frac{15377}{186624}\,e^{4}\right),
 \nonumber \\
  & \mathcal{L}_1 = \left(1+\frac{19}{36}\,e^{2}+\frac{1993}{5184}\,e^{4}\right)
\left(1-\frac{\iota^{2}}{2}+\frac{\iota^{4}}{24}\right)
- k_2\,\iota^{2}\left(6-5\iota^{2}\right)
\left(\frac{1}{12}-\frac{77}{864}\,e^{2}+\frac{361}{186624}\,e^{4}\right), \nonumber \\
 & \mathcal{L}_2 = \left(2+\frac{83}{18}\,e^{2}+\frac{2093}{288}\,e^{4}\right)
- \left(1+\frac{215}{72}\,e^{2}+\frac{93373}{15552}\,e^{4}\right)
\left(\iota^{2}-\frac{\iota^{4}}{3}\right) \nonumber \\
 & \qquad + k_2\,\iota^{2}\left[-1+\frac{67}{72}\,e^{2}+\frac{1709}{5184}\,e^{4}
+\left(\frac{13}{12}-\frac{611}{432}\,e^{2}-\frac{13175}{41472}\,e^{4}\right)\iota^{2}\right], \nonumber \\
 & \mathcal{K}_1 =\left(1+\frac{1}{36}\,e^{2}-\frac{23}{5184}\,e^{4}\right)
+ k_2\,\iota^{2}\left(\iota^{2}-3\right)
\left(\frac{1}{6}-\frac{113}{432}\,e^{2}+\frac{6733}{93312}\,e^{4}\right), \nonumber \\
 & \mathcal{K}_2 = \left(1+\frac{35}{36}\,e^{2}+\frac{1771}{1728}\,e^{4}\right)
\left(1-\frac{\iota^{2}}{2}+\frac{\iota^{4}}{24}\right)
- k_2\,\iota^{2}\left(6-5\iota^{2}\right)
\left(\frac{1}{12}+\frac{191}{864}\,e^{2}-\frac{829}{62208}\,e^{4}\right).
\label{eq:K's}
\end{align} 
\end{widetext}
%%%%%
%%%
Interestingly, in the above set of equations, there exists no term with odd powers of $\iota$. This supports the fact that Kerr has the symmetry about the equatorial plane. However, if we set $k_2=0$, the leading terms $\mathcal{E}_1$ and $\mathcal{K}_1$ do not depend on $\iota$, and $\mathcal{L}_1$ depends on it only through the projection $\cos\iota$. The remaining $\iota$-dependence, in $\mathcal{E}_2$, $\mathcal{L}_2$ and $\mathcal{K}_2$, comes from the spin and disappears for $S=0$. This shows that if we only introduce the spin and have not modified the dark matter distribution, there would still be effects from the inclination, which is expected. We now employ \ref{D1} to obtain the rate of change of eccentricity, semi-major axis, and inclination. They are written as:
\begin{widetext}
\begin{align}
\frac{\langle\dot e_{\rm dm}\rangle}{e}
&= \frac{k_1\beta\eta}{972\,(a/M)^{5/6}}
\Big[\,648+117e^{2}-\big(108-65e^{2}\big)k_2\iota^{2}\big(\iota^{2}-3\big)\Big]+\frac{k_1\beta\eta}{3888\,(a/M)^{7/3}}\frac{S}{M^{2}}
\Big\{-108\Big[24-12(1+2k_2)\iota^{2}
\nonumber\\
&\qquad \qquad+(1+20k_2)\iota^{4}\Big]
+e^{2}\Big[-2664+6(222-373k_2)\iota^{2}+(1865k_2-111)\iota^{4}\Big]\Big\},
\nonumber\\[10pt]
\frac{\langle\dot a_{\rm dm}\rangle}{a}
&= -\frac{k_1\beta\eta}{46656\,(a/M)^{5/6}}
\Big\{15552\Big[6+k_2\iota^{2}(\iota^{2}-3)\Big]
+216e^{2}\Big[156+7k_2\iota^{2}(\iota^{2}-3)\Big]
+5e^{4}\Big[4410+401k_2\iota^{2}(\iota^{2}-3)\Big]\Big\}
\nonumber\\
&\quad+\frac{k_1\beta\eta}{15552\,(a/M)^{7/3}}\frac{S}{M^{2}}
\Big\{5184\Big[24-12(1+k_2)\iota^{2}+(1+10k_2)\iota^{4}\Big]
+576e^{2}\Big[480-6(40+23k_2)\iota^{2}+5(4+23k_2)\iota^{4}\Big]
\nonumber\\
&\hspace{9.5em}+e^{4}\Big[465480-12(19395+7729k_2)\iota^{2}+5(3879+15458k_2)\iota^{4}\Big]\Big\},
\nonumber\\[10pt]
\frac{\langle\dot\iota_{\rm dm}\rangle}{\iota}
&=\frac{k_1\beta\eta}{(a/M)^{7/3}}\frac{S}{M^{2}}
\Big[1+\frac{17}{8}e^{2}+\frac{38117}{15552}e^{4}
-\Big(\frac{2+3k_2}{12}+\frac{51+43k_2}{144}e^{2}
+\frac{152468+21903k_2}{373248}e^{4}\Big)\iota^{2}\Big]
\label{eq:evolution}
\end{align}
\end{widetext}
Note that in the $S=0$ limit, the rate of change of inclination vanishes and $\iota$ is constant over the evolution. One can realize this by studying \ref{eq:dissip}, where if we set $S=0$, the orientation of the plane does not change. Therefore, the value of $\iota$ remains constant in time. 

For the limit $S=\iota=k_2=0$, we observe that
\begin{equation}
    \dfrac{\langle\dot{e}_{\rm dm}\rangle}{e}=\dfrac{k_1 \beta \eta (72+13 e^2)}{108 (a/M)^{5/6}},
\end{equation}
which is a positive quantity and suggests an eccentrification ($\dot{e}>0$), rather than a circularization ($\dot{e}<0$) \cite{Becker:2021ivq}. However, if we consider these $\iota$-dependent effects, one may ask whether the dark matter effects may cause circularization instead of eccentrification. We can take the spin-independent effect, and arrive at:
%%%
\begin{align}
\frac{\langle\dot e_{\rm dm}\rangle}{e}\bigg|_{S=0}
=\frac{k_1\beta\eta}{972\,(a/M)^{5/6}}
\Big[\,648+117e^{2}+ \nonumber \\
\big(108-65e^{2}\big)\,k_2\,\iota^{2}\big(3-\iota^{2}\big)\Big].
\end{align}
%%%
Interestingly, within the small $\iota$ approximation, as we are attempting in the present case, the above expression is always positive and indicates eccentrification. 

The other point that we should now touch upon is the rate of change of $\iota$, which is entirely stemming from the spin of the Kerr BH. The anisotropy enters only through the $k_2$ terms of the coefficients. However, the leading term (for $\iota\to0$ and $e\to0$) does not depend on $k_2$ at all. The rate follows from $\cos\iota=L_z/\sqrt{K}$ [\ref{D1}],
\begin{equation}
  \dot{\iota}=-\frac{1}{\sin\iota}\left[\frac{\dot{L}_z}{\sqrt{K}}-\frac{L_z\,\dot{K}}{2K^{3/2}}\right], 
\end{equation}
and since the fluxes are known only up to $\mathcal{O}(\iota^4)$, $\dot{\iota}/\iota$ is reliable only up to $\mathcal{O}(\iota^2)$.
%
%%%%
\subsection{Numerical results}
%%% 
 In this subsection, we provide a couple of numerical examples where we carry out a binary evolution under GW and dark matter effects. To include these dissipative effects, we use the relations given in \ref{eq:ELKdotdm} to 
\ref{eq:K's}, and \ref{DM 18}. The total rates are the sum of the GW and dark matter contributions, $\dot{{E}}=\dot{{E}}_{\rm GW}+\dot{{E}}_{\rm dm}$, and similarly for $L_z$ and $K$. These are converted into evolution equations for $(a,e,\iota)$ using \ref{D1}. With this, we obtain $e(t)$, $a(t)$, and $\iota(t)$, which will be employed to evolve \ref{eq:rad_theta_time}, and we finally get $r(t)$, $\theta(t)$, and $\phi(t)$. Later, we use these expressions to obtain $x(t)$, $y(t)$, and $z(t)$, and then the quadrupole moments. Finally, using these relations, we obtain the gravitational waveform by using the quadrupole formula \cite{Maggiore2007}. 
%%%%
\subsubsection{Evolution of orbital elements}
\begin{figure*}[t]
  \centering
  \begin{minipage}[t]{0.33\textwidth}
    \centering
    \includegraphics[width=\linewidth]{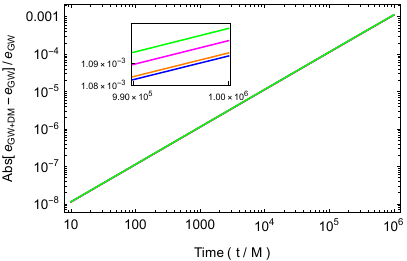}
  \end{minipage}\hfill
  \begin{minipage}[t]{0.33\textwidth}
    \centering
    \includegraphics[width=\linewidth]{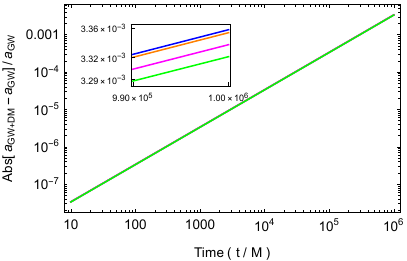}
  \end{minipage}\hfill
  \begin{minipage}[t]{0.33\textwidth}
    \centering
    \includegraphics[width=\linewidth]{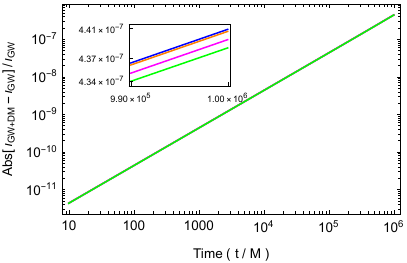}
  \end{minipage}

  \vspace{4mm}

  \begin{minipage}[t]{0.33\textwidth}
    \centering
    \includegraphics[width=\linewidth]{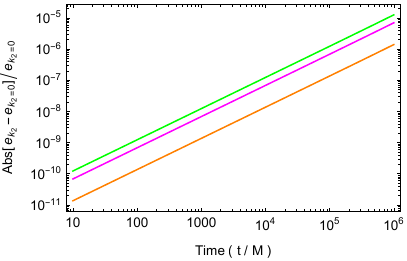}
  \end{minipage}\hfill
  \begin{minipage}[t]{0.33\textwidth}
    \centering
    \includegraphics[width=\linewidth]{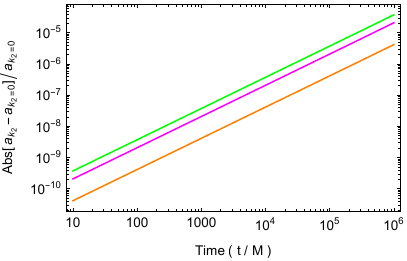}
  \end{minipage}\hfill
  \begin{minipage}[t]{0.33\textwidth}
    \centering
    \includegraphics[width=\linewidth]{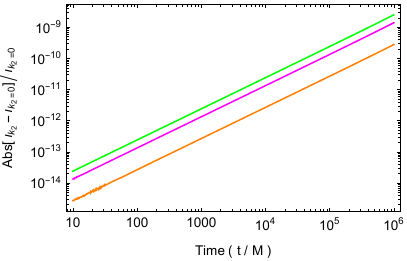}
  \end{minipage}
\caption{Effect of dark matter on the orbital elements. Top row: relative difference $|X_{\rm GW+dm}-X_{\rm GW}|/X_{\rm GW}$ for $X=(e, a, \iota)$, for $k_2=0$ (blue), $0.1$ (orange), $0.5$ (magenta) and $0.9$ (green); the insets zoom on the late-time region. Bottom row: effect of the anisotropy alone, i.e. the difference from the $k_2=0$ curve, for $k_2=0.1$ (orange), $0.5$ (magenta) and $0.9$ (green). The initial conditions are $(a_{\rm in},e_{\rm in},\iota_{\rm in},S,\eta)=(50M,0.1,\pi/20,0.09M^2,10^{-3})$ and $k_1=10^{-6}$.}
  \label{fig:orb_dm}
\end{figure*}

%%%%
We first compare the evolution of $e(t)$, $a(t)$ and $\iota(t)$ with and without dark matter. In \ref{fig:orb_dm}, we plot the relative differences of these orbital elements as functions of time, for several values of $k_2$ and with $k_1=10^{-6}$. For the eccentricity, the dark matter changes $e$ by $\mathcal{O}(10^{-3})$ at $t=10^6M$ relative to the evolution without dark matter, and increasing $k_2$ gives a larger departure. The anisotropy alone changes $e$ by at most $\mathcal{O}(10^{-5})$ relative to $k_2=0$, i.e.\ by about one percent of the dark matter effect. For the semi-major axis, the dark matter effect is also $\mathcal{O}(10^{-3})$. However, here increasing $k_2$ reduces the departure, because the lower density seen by an inclined orbit weakens the drag. The inclination is affected far less, by $\mathcal{O}(10^{-7})$, and the anisotropy changes it only at the level of $\mathcal{O}(10^{-9})$. This is expected as $k_2$ enters $\dot\iota/\iota$ only at relative order $\iota^2$.
%%%%
%%%%
\subsubsection{Dark matter imprints on the waveform}
\begin{figure*}[!t]
\centering
\includegraphics[width=\textwidth]{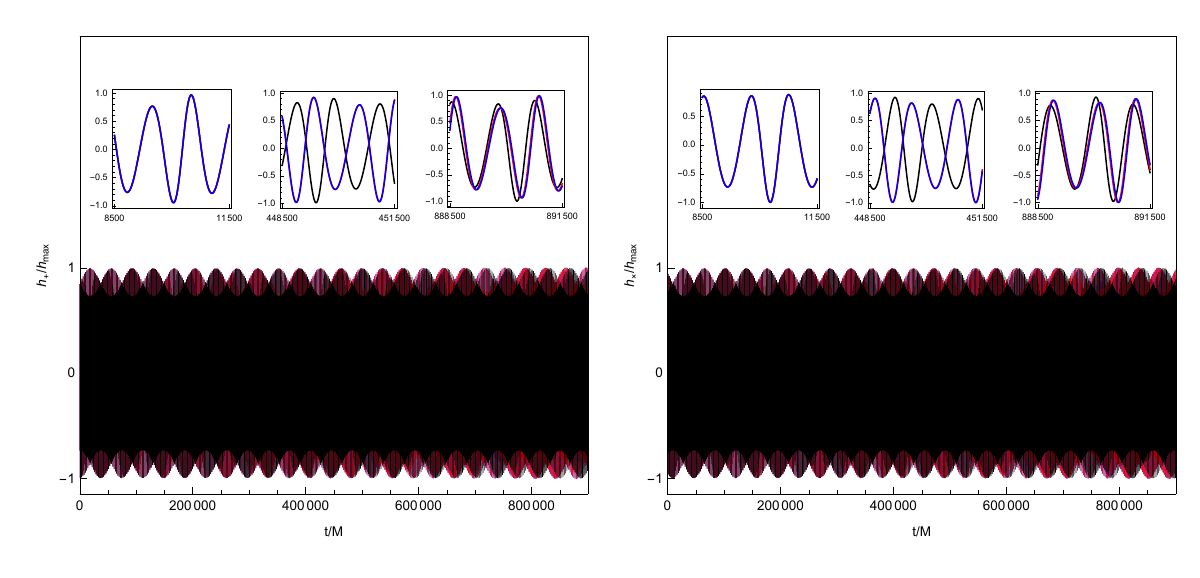}
\caption{In the above figure, we have shown plus (left) and cross (right) polarizations, normalized to the maximum of the waveform without dark matter, as a function of $t/M$. The coloring has the following meanings -- black: without dark matter; red: with the dark matter spike ($k_1=10^{-6}$, $k_2=0$); blue: with the dark matter spike ($k_1=10^{-6}$, $k_2=0.9$). The three insets in each panel magnify windows of width $3\times10^{3}\,M$ centred at $t=10^{4}\,M$, $4.5\times10^{5}\,M$ and $8.9\times10^{5}\,M$. The initial conditions are the same as \ref{fig:orb_dm}.}
\label{fig:wave_full}
\end{figure*}

%%%%
In \ref{fig:wave_full}, we compare the plus (left panel) and cross (right panel) polarizations, computed with the quadrupole formula along the evolved orbit, for three cases: the evolution without dark matter, the evolution with the dark matter spike and $k_2=0$, and the evolution with the dark matter spike and $k_2=0.9$. All waveforms are normalized to the maximum of the waveform without dark matter. The dark matter coupling is $k_1=10^{-6}$, and other initial conditions are $(a_{\rm in},e_{\rm in},\iota_{\rm in},S,\eta)=(50M,0.1,\pi/20,0.09M^2,10^{-3})$.

Over the full evolution, the three waveforms have the same overall amplitude profile, since the dark matter correction to the orbital elements is small (\ref{fig:orb_dm}). The difference appears in the phase. To show this, the three insets zoom into windows around $t=10^{4}M$, $4.5\times10^{5}M$ and $8.9\times10^{5}M$. In the first inset, the three curves are almost indistinguishable. In the later insets, the curves with the dark matter spike shift relative to the curve without dark matter, and this shift grows with time. The curves with $k_2=0$ and $k_2=0.9$ separate at late times, which shows that the anisotropy, controlled by $k_2$, affects the waveform phase even though its effect on the orbital elements is much smaller (\ref{fig:orb_dm}). We quantify this accumulated phase difference in the next subsection.

%%%%
\subsubsection{Dephasing of the gravitational-wave signal}
%%%%
%%%%%%%\textit{}
  \begin{figure*}[t]
  \centering
  \includegraphics[width=\textwidth]{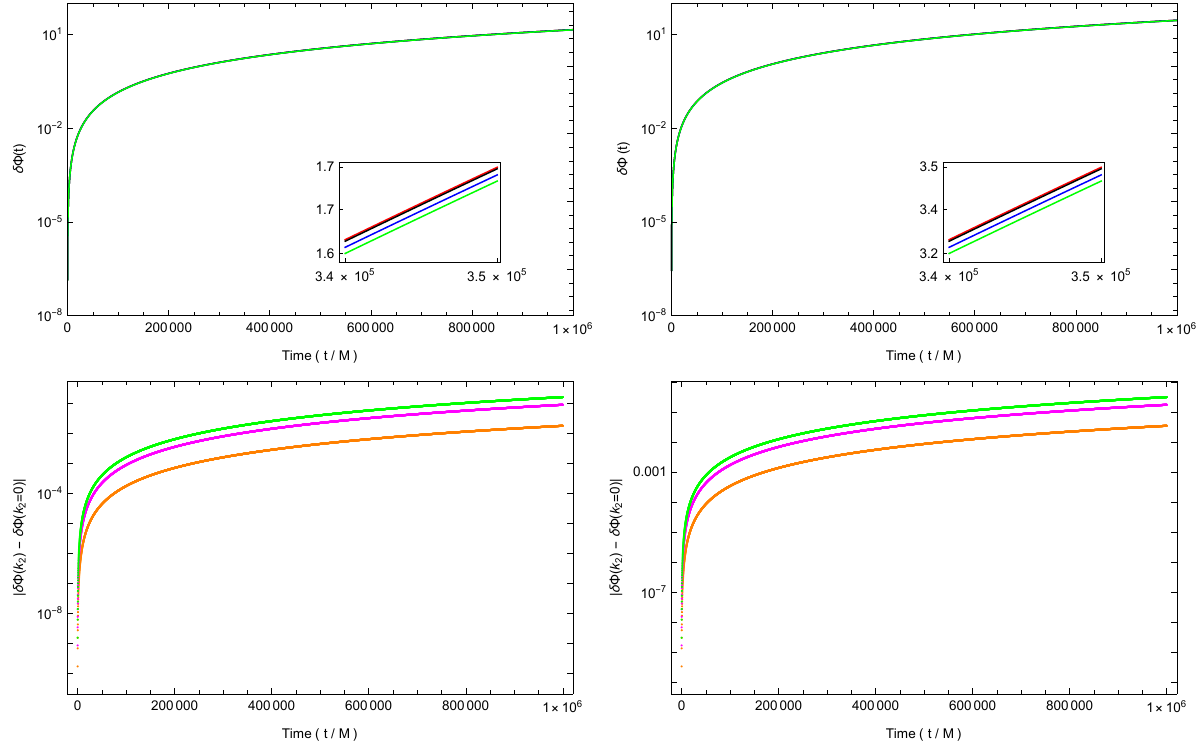}
  \caption{Dephasing $\delta\Phi=\Phi_{\rm gw+dm}-\Phi_{\rm gw}$ as a function of $t/M$ for the $(2,0,0)$ mode (left column) and the $(2,1,1)$ mode (right column), for $k_1=10^{-6}$. Top row: $\delta\Phi$ for $k_2=0$ (red), $0.1$ (black), $0.5$ (blue) and $0.9$ (green). The inset magnifies the window $t=3.4\times10^{5}\,M$ to $3.5\times10^{5}\,M$, where the curves for different $k_2$ separate. Bottom row: the effect of $k_2$ alone, $|\delta\Phi(k_2)-\delta\Phi(k_2=0)|$, for $k_2=0.1$ (orange), $0.5$ (magenta) and $0.9$ (green). In all panels the mass ratio is $\eta=10^{-3}$ and the initial conditions are $(a_{\rm in},e_{\rm in},\iota_{\rm in},S)=(50M,\,0.1,\,\pi/20,\,0.09M^2)$. The $(2,1,1)$ mode gives the larger dephasing.}
  \label{fig:wide_3}
\end{figure*}
%%%%
The other aspect that we illustrate is the phase of the emitted gravitational waves (more details are given in Appendix \ref{sec:dephasing}). To quantify the dark matter contribution, we compute the phase difference
%%%%
\begin{equation}
    \delta \Phi = \Phi_{\mathrm{gw+dm}} - \Phi_{\mathrm{gw}},
    \label{relativephase}
\end{equation}
%%%%
where $\Phi_{\mathrm{gw+dm}}$ is the GW phase evolution including both dark matter and gravitational radiation, and $\Phi_{\mathrm{gw}}$ is the phase evolution driven solely by the GW radiation reaction. Both are evaluated for the same mode $(m,k,n)$, at the same time $t$, and for the same initial $(a,e,\iota)$. 

\ref{fig:wide_3} shows the dephasing for $\eta=10^{-3}$, for the $(2,0,0)$ mode (left column) and the $(2,1,1)$ mode (right column). The top row shows $\delta\Phi$ against $t/M$ for $k_2=0,\,0.1,\,0.5$ and $0.9$, on a logarithmic vertical axis. As expected, the accumulated dephasing grows steadily as the binary evolves. At $t=10^{6}M$ and for $k_2=0.5$ it reaches $14.2$ rad for the $(2,0,0)$ mode and $28.4$ rad for the $(2,1,1)$ mode. The $(2,1,1)$ mode gives the larger dephasing, by a factor of $2.0$. This is expected: the phase is $\Phi_{mkn}=n\Phi_r+k\Phi_\theta+m\Phi_\phi$, and the three angular velocities are equal at leading order, so the phase of the $(2,1,1)$ mode is about twice that of the $(2,0,0)$ mode.

In the top row, the curves for different $k_2$ lie close together on this scale, but they separate at late times. The inset magnifies the window $t=3.4\times10^{5}M$ to $3.5\times10^{5}M$, where the curves separate, and $\delta\Phi$ decreases as $k_2$ increases. The bottom row shows the effect of $k_2$ alone, $|\delta\Phi(k_2)-\delta\Phi(k_2=0)|$. The signed difference is negative, so $k_2$ reduces the dephasing, and its magnitude grows with time and with $k_2$; the maximum dephasing for each $k_2$ is given in \ref{tab:scan}. The anisotropy therefore leaves a secular imprint on the phase, even though its effect on the orbital elements is much smaller than that of the radial spike.

\begin{figure*}[t]
\includegraphics[width=\textwidth]{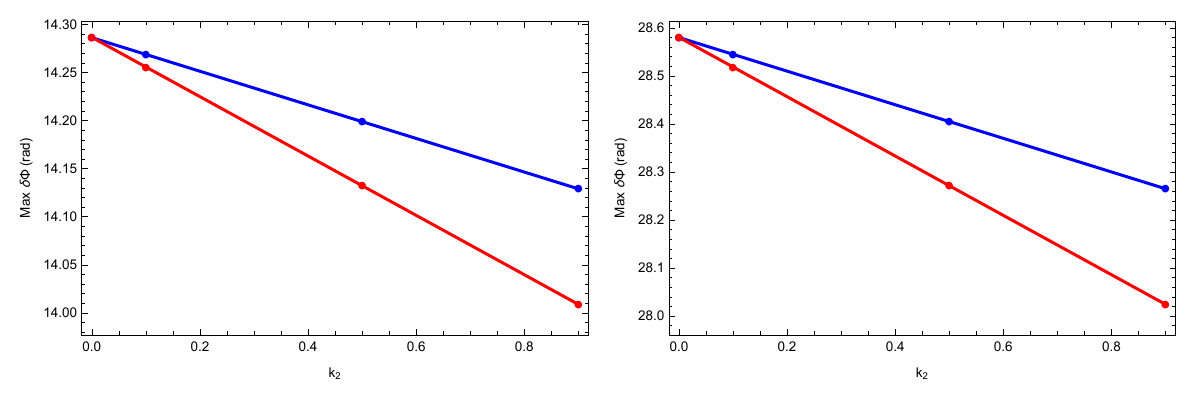}
\caption{In the above, we have shown maximum dephasing as a function of $k_2$ for the $(2,0,0)$ (left) and $(2,1,1)$ (right) modes, for $\iota_{\rm in}=\pi/20$ (blue) and $\pi/15$ (red). The reduction with $k_2$ grows with the initial inclination. Other parameters: $a_{\rm in}=50M$, $\eta=10^{-3}$, $k_1=10^{-6}$, $e_{\rm in}=0.1$ and $S=0.09M^2$.}
\label{fig:incl}
\end{figure*}

In \ref{tab:scan} we give the maximum dephasing of the $(2,0,0)$ and $(2,1,1)$ modes over $t\le10^{6}M$, for the mass ratios $\eta=10^{-3}$ and $5\times10^{-4}$ and the initial semi-major axes $a_{\rm in}=30M$ and $50M$. For $a_{\rm in}=50M$ the dephasing increases with $\eta$: halving $\eta$ from $10^{-3}$ to $5\times10^{-4}$ reduces it by a factor of about $1.8$. The dephasing decreases with $a_{\rm in}$, and the $(2,1,1)$ mode is larger than the $(2,0,0)$ mode by a factor of $2.0$ for $a_{\rm in}=50M$. Increasing $k_2$ lowers the dephasing, because for $k_2>0$ the density is depleted away from the equatorial plane, and an inclined orbit spends part of its time there. The effect therefore depends on the inclination. To highlight the effect of $k_2$, we show the maximum dephasing in \ref{fig:incl} as a function of $k_2$ for $\iota_{\rm in}=\pi/20$ and $\pi/15$, with $a_{\rm in}=50M$ and $\eta=10^{-3}$. The value at $k_2=0$ is nearly independent of $\iota_{\rm in}$, while the reduction at $k_2=0.9$ increases as we increase $\iota$ from $\pi/20$ to $\pi/15$. The $(2,1,1)$ mode shows a similar trend as the $(2,0,0)$ mode.
%%%
\begin{table}[t]
\caption{Maximum dephasing $\max\delta\Phi$ (rad) over $t\le10^{6}M$ for the $(2,0,0)$ and $(2,1,1)$ modes, for different initial semi-major axes $a_{\rm in}$, mass ratio $\eta$ and values of $k_2$, with $k_1=10^{-6}$, $e_{\rm in}=0.1$, $\iota_{\rm in}=\pi/20$ and $S=0.09M^2$.}
\label{tab:scan}
\begin{tabular}{cccccc}
\hline\hline
 &  & \multicolumn{2}{c}{$\eta=10^{-3}$} & \multicolumn{2}{c}{$\eta=5\times10^{-4}$} \\
\cline{3-4}\cline{5-6}
$a_{\rm in}/M$ & $k_2$ & $(2,0,0)$ & $(2,1,1)$ & $(2,0,0)$ & $(2,1,1)$ \\
\hline
50 & 0   &14.2862 & 28.5795 & 7.8280 & 15.6600 \\
   & 0.1 & 14.2687 & 28.5446 & 7.8185 & 15.6408 \\
   & 0.5 & 14.1988 & 28.4047 & 7.7802 & 15.5643 \\
   & 0.9 & 14.1289 & 28.2649 & 7.7419 & 15.4877 \\
\hline
30 & 0   & 48.9399 & 97.9334 & 26.2616 & 52.5516 \\
   & 0.1 & 48.8799 & 97.8134 & 26.2295 & 52.4874 \\
   & 0.5 & 48.6401 & 97.3336 & 26.1010 & 52.2303 \\
   & 0.9 & 48.4003 & 96.8538 & 25.9726 & 51.9733 \\
\hline\hline
\end{tabular}
\end{table}
%
%
%%%%%%%
%%%%%%%
%%%%%
\section{Discussion}
\label{sec: Discussion}
%%%%%
In the present article, we have studied how a dark matter spike in the vicinity of the slowly rotating central object affects the emitted GW during a binary evolution. Due to the rotation of the central object, the dark matter distribution is expected to become anisotropic and therefore, to pick up a $\theta$-dependence. Our primary motivation is twofold: first is to develop a framework to add these off-equatorial $\theta$-dependent effects into the emitted waveform with dark matter dissipation, and second, to estimate the dephasing contributed by the spike. The basic framework is discussed in \sect{sec: Basic framework} \& \sect{sec:Evolution}, whereas the dark matter model, the resulting evolution of the orbit, and the dephasing calculations are in \sect{sec:Numerical Result}. 

Once the central object is rotating, the dark matter distribution around it is expected to be modified. Within the present scheme, the central object is slowly rotating, so we keep only the leading angular deviation from spherical symmetry. The distribution, as shown in \ref{Darkmatterprofile}, is quadratic in $\cos\theta$, $\bar{\rho}\propto 1-k_2\cos^2\theta$ to leading order, where we introduce the parameter $k_2$ to control how it varies on both sides of the equator. On the other hand, radial distribution is controlled by $k_1$ which actually combines $\bar{\rho}_{\rm sp}$, $r_{\rm sp}$, and $\alpha$. Throughout the paper, we use $\alpha=7/3$, which makes the radial dependence fall as $r^{-7/3}$. In addition, we kept $k_1=10^{-6}$ fixed for the entire paper. In \ref{Darkmatterprofile}, we have shown a typical dark matter distribution for $k_1=10^{-6}$ and \(k_2 = 1/2\). It is significantly dense at lower \(r\) values and falls off as \(r^{-7/3}\) as \(r\) increases, and as we go away from the equatorial plane, the density falls off. Within such an environment, the dynamics is then studied, and both the polarization and phase shift have been calculated. Even if the distribution depends on both $r$ and $\theta$, it is written in a separable form. While it may not be necessary for a complete numerical calculation, as we are interested in obtaining the analytical expressions, the separability is assumed explicitly. This also makes the dissipative calculations relatively simpler, as shown in \sect{sec:Evolution}. In order to obtain these relations, we employ the technique used by Ryan in Ref. \cite{PhysRevD.53.3064}. Within this approach, we obtain the orbit-averaged dissipation of the energy, angular momentum and Carter constant as analytical expansions up to $\mathcal{O}(e^4,\iota^4,S)$, which we then use to evolve the orbit.

In \ref{fig:wave_full}, we compare the gravitational waveforms emitted without dark matter, with a purely radial spike ($k_2=0$), and with an anisotropic spike ($k_2=0.9$). The waveforms separate progressively, so that the difference accumulates over many orbital cycles. \ref{fig:wide_3} shows how the dephasing grows with time and how it depends on $k_2$. The radial spike sets the overall size of the dephasing, which is larger for a larger mass ratio, for a smaller initial semi-major axis, and for the $(2,1,1)$ mode than for the $(2,0,0)$ mode (\ref{tab:scan}). The anisotropy of the spike imprints itself on the phase. Increasing $k_2$ reduces the dephasing at all times, because an inclined orbit samples a lower dark matter density. The reduction grows with the initial inclination (\ref{fig:incl}), as expected from the orbit-averaged density, and it vanishes only for $k_2=0$ or for an equatorial orbit. Being secular, it builds up over the whole evolution. The plots show the qualitative feature that the anisotropy introduces a secular phase shift in the emitted GWs.  

Our results indicate that the angular structure of the dark matter environment cannot be ignored for generic binaries. For an orbit inclined with respect to the equatorial plane, the anisotropy lowers the density that the companion experiences. At a fixed equatorial density, an equatorial orbit has a larger dephasing than an inclined one, so using the equatorial value for a generic orbit overestimates the dephasing. The reduction accumulates with time and grows with the inclination. Hence, the dephasing depends on the orientation of the orbit, which a spherical spike cannot capture. We stress, however, that our model is simplified: the angular profile is a toy model, the dark matter drag is Newtonian, and the expansion in the eccentricity and the inclination restricts our results to small eccentricities and inclinations. A more complete treatment is left for future work.
%%%%%%%%%%%%%%%%%%%%%%%%%%%%%%
\section*{Acknowledgments}
%%%%%%%%%%%%%%%%%%%%%%%%%%%%%%%%%%%%
S. Maity and S. Mukherjee are thankful to the Inspire Faculty Grant (DST/INSPIRE/04/2020/001332) from DST, Govt. of India. In addition, S. Mukherjee is also thankful to the Prime Minister Early Career Research Grant (ANRF/ECRG/2024/004108/PMS) by ANRF, Govt. of India, and the New Faculty Seed Grant (NFSG/PIL/2023/P3794) provided by BITS Pilani (Pilani), India, for financial support. He (S. Mukherjee) is also grateful to the Visiting Associateship Program at IUCAA, Pune, for academic visits where a part of this work was carried out. R. L. acknowledges ANRF for the financial support through ARG (grant no. ANRF/ARG/2025/005140/PS). Finally, S. Mukherjee and R. L. acknowledge the LIGO Science Workshop, organized by the International Centre for Theoretical
Sciences (ICTS) from October 27-28, 2023, where this project
was originally conceived, and the Workshop on High Energy Physics Phenomenology (WHEPP-2025), held at IIT Hyderabad from December 14-20, 2025, where a part of this work was carried out.
%%%%%%%%%%%%%%%%%%%%%%%%%%%%%%%%%%%%
\appendix 
%%%%%%%%%
\section{The equations of motion}
\label{appendix:EOM}
%%%%%
Here we introduce the orbit equations that we have employed in the present paper. We start with the following expression for the Lagrangian \cite{PhysRevD.53.3064}
\begin{eqnarray}
    \mathcal{L}=\dfrac{m}{2}\Big(\dot{r}^2+r^2 \dot{\theta}^2+r^2 \sin^2\theta \dot{\phi}^2\Big)+ \dfrac{mM}{r}-\dfrac{2mS \sin^2\theta}{r}\dot{\phi}, \nonumber \\
\end{eqnarray}
%%%
which is written up to the linear order in black hole's spin. It admits three constants of motion, namely, energy ($\mathcal{E}$), orbital angular momentum ($L_z$) and Carter constant (K) given as
%%%%
\begin{eqnarray}
  &&  \mathcal{E}=\dfrac{m}{2}\Big(\dot{r}^2+r^2 \dot{\theta}^2+r^2 \sin^2\theta \dot{\phi}^2\Big)-M m/r,\nonumber \\
   && L_{z}=m r^2 \sin^2 \theta \dot{\phi}-2 m S\sin^2\theta/r, \nonumber \\
   && K=m^2 r^4(\dot{\theta}^2+\sin^2\theta \dot{\phi}^2)-4m^2 S r \sin^2\theta \dot{\phi}.
   \label{eq:EOM}
\end{eqnarray}
%%%%
we can now rewrite the above set of equations into the following form:
%%%
\begin{eqnarray}
&&m^2\Big(\dfrac{dr}{dt}\Big)^2=2 m \mathcal{E}+\dfrac{2Mm^2}{r}-\dfrac{K}{r^2}-\dfrac{4L_{z}mS}{ r^3}+\mathcal{O}(S^2),\nonumber \\
&&m^2\Big(\dfrac{d\theta}{dt}\Big)^2=\dfrac{K-L^2_z \csc^2\theta}{r^4}+\mathcal{O}(S^2), \nonumber \\
&& m\dfrac{d\phi}{dt}=\dfrac{L_z\csc^2\theta}{r^2}+\dfrac{2mS}{r^3}.
\label{eq:rad_theta_time}
\end{eqnarray}
%%%
The above expressions dictate how the radial and angular coordinates are changing with time. Once we solve the above, we obtain $r(t)$, $\theta(t)$ and $\phi(t)$. 
%%%%%%%%%
\section{Dissipation due to dark matter: generic orbits}
\label{appendix:dissipation_dm}
%%%%%
For a test mass $m$ moving through a dark matter distribution, $\rho_{dm}(r,\theta)$, the dragging force, $\vec{F}_{\text{dm}}$, acts against particle's motion and is defined as
%%%%
\begin{eqnarray}
\vec{F}_{\text{dm}} &=& \frac{4 \pi m^2 \rho_{dm}(r, \theta) \ln \Big(\sqrt{M/m}\Big)}{v^2}(-\hat{v}), \nonumber \\
&=&-\dfrac{\beta \eta^2 \bar{\rho}_{dm}(r,\theta)}{v^2}\hat{v}=-F_{\text{dm}}\hat{v}, \nonumber \\
 \label{DM 3}
\end{eqnarray}
%%%%%
where $\beta=4 \pi \ln (\sqrt{M/m})$ \cite{Kavanagh_2020}, and $v$ is the velocity of the lighter object. In the last step, we introduce $\eta$, which is the mass ratio, $\eta=m/M$, and  $\bar{\rho}_{\text{dm}}(r,\theta)$ is normalized with the mass $M$, i.e., $M^2\rho_{\text{dm}}(r,\theta)=\bar{\rho}_{dm}(r,\theta)$ to make it dimensionless. Within the Newtonian approximation, $v$ can be expressed as: 
\begin{equation}
v = \sqrt{\dfrac{2}{m}\left({\mathcal{E}} + \frac{m M}{{r}}\right)},
 \label{DM 5}
\end{equation}
which is consistent with the Newtonian description of $\mathcal{E}$ \cite{PhysRevD.53.3064}.
%
%%%%
%%%%
In order to evaluate the rate of change of energy due to $F_{\rm dm}$, we rewrite \ref{DM 5} as
%%%
\begin{equation}
\mathcal{E} = \frac{1}{2} mv^2 - \frac{Mm}{r}
\label{A1}
\end{equation}
%%%%%
By using \ref{DM 3}, the dissipation due to $F_{\rm dm}$ is given by
%%%
\begin{equation}
\frac{d \mathcal{E}}{dt} = \vec{v} \cdot \vec{F}_{\rm dm}=-\dfrac{\beta \eta^2}{v} \bar{\rho}_{dm}(r,\theta)
\label{A5}
\end{equation}
%%%
Given that $\mathcal{E}$ and $E$ only differs by a constant term, we obtain
\begin{equation}
   \dot{E}=\dfrac{dE}{dt}=-\dfrac{\beta \eta^2}{v} \bar{\rho}_{dm}(r,\theta),
\end{equation}
Similarly, we can also obtain the rate of change of angular momentum in a Newtonian approach. Given the drag force is $\vec{F}_{\rm dm}$, the rate of change of angular momentum is given as
%%%
\begin{eqnarray}
   \dot{\vec{l}}=\vec{r}\times \vec{F}_{\rm dm}=-\dfrac{F_{\rm dm}}{mv}(m\vec{r} \times \vec{v})=-\dfrac{F_{\rm dm}}{mv}\vec{l}, 
   \label{eq:lvecdot}
\end{eqnarray}
%%%
where we introduce the mechanical angular momentum, $\vec{l}$ as $\vec{l}= m\vec{r} \times \vec{v}$, with $\vec{v}$ as the velocity vector. To emphasize, $l_{z}=m r^2 \sin^2\theta \dot{\phi}$, whereas, 
\begin{equation}
L_{z}=l_z-\dfrac{2mS \sin^2\theta}{r}.     
\end{equation}
As we are particularly interested in an axially-symmetric spacetime in which the rotation axis coincides with the z-axis, we have
%%%
\begin{eqnarray}
    \dot{l_z} = -\dfrac{F_{\rm dm}}{mv}{l_z}=-\dfrac{\beta \eta^2 \bar{\rho}(r,\theta)}{mv^3}{l_z} \nonumber \\
    =-\dfrac{\beta \eta^2 \bar{\rho}(r,\theta)}{mv^3}\Big(L_z+\dfrac{2mS \sin^2\theta}{r}\Big).
    \label{eq:Lzdot}
\end{eqnarray}
%%%%
The spin term in $L_z$ depends only on the position, which the drag does not change, so the drag contributes to $\dot{L}_z$ exactly as it does to $\dot{l}_z$. Hence, in an axially-symmetric spacetime in which the rotation axis coincides with the $z$-axis, we have
%%%
\begin{eqnarray}
    \dot{L_z}=-\dfrac{\beta \eta^2 \bar{\rho}(r,\theta)}{mv^3}\Big(L_z+\dfrac{2mS \sin^2\theta}{r}\Big).
\end{eqnarray}
%%%%
We now use \ref{eq:lvecdot} to obtain the rate of change of total angular momentum $l^2=\vec{l}.\vec{l}$ as follows:
%%%
\begin{equation}
        \dfrac{dl^2}{dt}=-2\Big(\dfrac{F_{\rm dm}}{mv}\Big)l^2.
\end{equation}
%%%
With these relations obtained, we now attempt to obtain the dissipation of ${K}$.  Given we are only interested up to $\mathcal{O}(S)$, the relation between $K$ and $L^2$ reads as \cite{PhysRevD.53.3064}:
%%%%
\begin{equation}
    K=l^2-\dfrac{4}{r}mSL_{z}+\mathcal{O}(S^2).
    \label{eq:Carter}
\end{equation}
%%%%
Therefore, the rate of change of ${K}$ becomes:
%%%
\begin{eqnarray}
   \dot{K}= \dfrac{dK}{dt} &=&-2 \dfrac{F_{\rm dm}}{mv}l^2-\dfrac{4 m S}{r}\dot{L}_z.
\end{eqnarray}
%%%%%
In the above expression, we can use \ref{eq:Carter} and rewrite $l^2$ in terms of $K$, and obtain:
%%%
\begin{eqnarray}
   \dot{K}= \dfrac{dK}{dt} &=&-2 \dfrac{F_{\rm dm}}{mv}K-\dfrac{4 S F_{\rm dm}}{r v}{L}_z,
\end{eqnarray}
%%%
where we have ignored all the $\mathcal{O}(S^2)$ terms. For the equatorial plane, we have $K=L^2_{\rm z}$, and the above expression becomes:
%%%
\begin{eqnarray}
   \dot{K}= \dfrac{dK}{dt} &=&-2 \dfrac{F_{\rm dm}}{mv}L_z\Big(L_z+\dfrac{2mS}{r}\Big)=2 L_z \dot{L}_z, \nonumber \\
\end{eqnarray}
%%%
which matches the expected result. 
%%%%%
%%%%%
%%%
\section{Frequencies and total phase}
\label{sec:dephasing}
%%%
Given that the orbits are generic, the total phase is contributed by all three angles: radial, polar, and azimuthal. Following Ref. \cite{PhysRevD.53.3064}, the angular frequencies of the radial, polar and azimuthal motion, to linear order in the spin, are:
\begin{align}
  \Omega_r &=\Big(\dfrac{M}{a^3}\Big)^{1/2}+\dfrac{3M^2 \cos\iota}{a^3(1-e^2)^{3/2}}(S/M^2), \nonumber \\
  \Omega_\theta &=\Big(\dfrac{M}{a^3}\Big)^{1/2}-\dfrac{3 M^2 \cos\iota}{a^3(1-e^2)^{3/2}}(S/M^2), \nonumber \\
  \Omega_\phi &=\Big(\dfrac{M}{a^3}\Big)^{1/2}+\dfrac{(2-3 \cos\iota)M^2}{a^3(1-e^2)^{3/2}}(S/M^2).
\end{align}
The total phase, $\Phi_{mkn}(t)$, is a sum of radial $\Phi_{r}(t)$, polar, $\Phi_{\theta}(t)$, and azimuthal phases, $\Phi_{\phi}(t)$, and given by:
\begin{equation}
 \Phi_{mkn}(t)=n \Phi_{r}(t)+k \Phi_{\theta}(t)+m \Phi_{\phi}(t), 
 \vspace{5mm}
\end{equation}
where $(m,k,n)$ represents different modes. The radial, polar, and azimuthal phases, written in terms of their respective angular velocities, are given as follows:
\begin{equation}
\Phi_r(t)=\int_{0}^t\Omega_r\,dt,~
\Phi_{\theta}(t)=\int_{0}^t\Omega_{\theta}\,dt,~
\Phi_{\phi}(t)=\int_{0}^t\Omega_{\phi}\,dt.
\end{equation}
%%
%%%%
\section{Dissipation due to GW radiation}
\label{sec:GW_rad}
%%%
Following Ref. \cite{PhysRevD.53.3064} (we set $\psi_0=0$ in Eq.~(14) of that reference), we now express the rate of change of energy, momentum, and Carter constant due to GW radiation up to $\mathcal{O}(e^4,\iota^4,S)$ terms as follows:
%%% 
%%%
\begin{widetext}
\begin{eqnarray}
 &&   \langle\dot{E}_{\text{GW}}\rangle = -\frac{32}{5} \left( \frac{m^2}{M^2} \right) \left( \frac{M}{a} \right)^5 
\left( \frac{1}{1 - e^2} \right)^{7/2} \Bigg[ \left(1 + \dfrac{73}{24}e^2 + \dfrac{37}{96}e^4\right) 
- \left( \dfrac{S}{M^2} \right) \left( \dfrac{M}{a(1 - e^2)} \right)^{3/2} \left(1-\dfrac{\iota^2}{2}+\dfrac{\iota^4}{24}\right)\nonumber \\
&& \hspace{10cm}\left( \dfrac{73}{12} + \dfrac{1211}{24}e^2 + \dfrac{3143}{96}e^4 \right) \Bigg], \nonumber \\
&&\langle\dot{L_z}_{\text{GW}}\rangle = -\frac{32}{5} \left( \frac{m^2}{M} \right) 
\left( \frac{M}{a} \right)^{7/2} \left( \dfrac{1}{1 - e^2} \right)^2 \Bigg[ 
\left( 1 + \dfrac{7}{8}e^2 \right) \left(1-\dfrac{\iota^2}{2}+\dfrac{\iota^4}{24} \right)
+ \left( \dfrac{S}{M^2} \right) \left( \dfrac{M}{a(1 - e^2)} \right)^{3/2} \nonumber \\
&& \hspace{1cm} \Big\{\left( \dfrac{61}{24} + \dfrac{63}{8}e^2 + \dfrac{95}{64}e^4 \right)
- \left(1-\iota^2+\dfrac{\iota^4}{3} \right)\left(\dfrac{61}{8} + \dfrac{109}{4}e^2 + \dfrac{293}{64}e^4 \right)-\left(\iota^2-\dfrac{\iota^4}{3} \right)\Bigl(\dfrac{5}{4}e^2+\dfrac{13}{16}e^4 \Bigr) \Big\}\Bigg], \nonumber \\
&&\langle\dot{K}_{\text{GW}}\rangle = -\dfrac{64}{5} {m^3} \left( \dfrac{M}{a} \right)^3 
\left( \dfrac{1}{1 - e^2} \right)^{3/2} \Bigg[ \left( 1 + \dfrac{7}{8}e^2 \right) 
- \left( \dfrac{S}{M^2} \right) \left( \dfrac{M}{a(1 - e^2)} \right)^{3/2} \left(1-\dfrac{\iota^2}{2}+\dfrac{\iota^4}{24} \right)\Bigl( \dfrac{97}{12} + {22}e^2 + \dfrac{99}{32}e^4 \Bigr) \Bigg]. \nonumber \\
\label{DM 18}
\end{eqnarray}
\end{widetext}
%%
%%%%%%%%%%%%%%%%%%%%%%%%%%%%%%%%%%%%
\bibliography{references}
\end{document}